\documentclass{article}
\usepackage{natbib}
\usepackage{icrctc07}

\title{Cosmic ray acceleration at modified shocks}
\shorttitle{Particle acceleration at modified shocks}
\authors{A. Meli$^{1}$, P. L. Biermann$^{2,3,4}$, S. Dimitrakoudis$^{1}$.}
\shortauthors{A. Meli et al.}
\afiliations{$^1$Department of Physics, National University of Athens, 15783 Panepistimiopolis Zografos, Greece\\ $^2$ Max Planck Institut for Radioastronomy, Auf dem Huegel 69, Bonn Germany\\ $^3$ Department of Physics and Astronomy, University of Bonn, Germany\\ 
$^4$ Department of Physics and Astronomy, University of Alabama, Tuscaloosa, Alabama 35487, USA}
\email{ameli@phys.uoa.gr}

\abstract{The non-linear back reaction of accelerated cosmic rays at the
shock fronts, leads to the formation of a smooth precursor with a length scale
corresponding to the diffusive scale of the energetic particles. 
Past works claimed that shocklets could be created
in the precursor region of a specific shock width, which might energize few thermal
particles to sufficient acceleration and furthermore this precursor region may
act as confining large angle scatterer for very high energy cosmic rays. On the other hand, 
it has been shown that the smoothing of the shock front could lower the acceleration efficiency.
These controversies motivated us to investigate numerically by Monte Carlo simulations the 
particle acceleration efficiency in oblique modified shocks. The results 
show flatter spectra compared to the spectra of the pressumed sharp discontinuity shock fronts. 
The findings are in accordance with theoretical predictions, since the scattering inside the precursor confines high energy particles to further scattering, resulting in higher energies making the whole acceleration process more efficient.}

\begin{document}
\maketitle


\section{Introduction}

It has been shown theoretically 30 years ago that the first order Fermi  acceleration, 
namely also diffusive shock acceleration (e.g. \cite{Bell78a}, \cite{Bell78b}), is a plausible 
mechanism for cosmic rays to gain statistically an amount of energy by crossing a shock front
-formed in a super-Alfv{\'e}nic plasma flow- in consecutive cycles,  while scattering off
the irregularities of the magnetic field present in the media. The energy
distribution of these particles follows a power law, which generally 
is in accordance with the actual measurable value of the spectral
index of the observed cosmic ray spectrum, at the top of the atmosphere.
Over the years, in the studies of diffusive shock acceleration it is commonly considered 
that a shock front is an immaterial non-dimensional surface, although this is not true. 
It quickly became evident to the researchers that the test-particle approximation, 
where the particles do not interact with the shock, could not entirely hold and  
non-linear effects could take the leading part into the dynamics of the shock acceleration mechanism. 
In other words, if a reasonable amount of energy was transfered to the accelerated particles (i.e. $>10\%$), 
they could dynamically play an important role in the shock process itself  
by interacting with the shock eventually modifying it.
Moreover, the modified shock which will consequently have a finite width (a precursor), will 
have a length scale which will correspond to the diffusive length scale of the energetic particles. \\
Specifically, work of \cite{Berezhko86}, \cite{Zank87}, \cite{Zank89} and \cite{Zank90} 
questioned the simplification of the test-particle approximation, which consequently 
considered the shock as an immaterial surface into the acceleration mechanism process
and moreover showed that cosmic ray influenced shock structures are unstable, 
indicating that the turbulence present could be strong, requiring a full treatment
of a violently scattering medium as a result. This instability  present in the modified shock
structures, could be of a great importance and a numerical investigation following
the above theoretical works, could throw further light into the
exact mechanism, particularly for oblique shocks. Concerning the latter, work of \cite{MeliBie06} has 
shown that highly non-relativistic oblique shocks could be very efficient cosmic ray accelerators under 
given circumstances of the particles diffusion \cite{Jokipii87}. 
Therefore, the results presented here could have further 
implications on the efficiency of these shocks, provided there is no immaterial shock surface as assumed before, but rather the conditions which are described in this work.
Summarizing, in this work we aim to test and compare past theoretical claims with numerical simulations
and investigate changes in the acceleration efficiency, spectral indexes and acceleration rates, simulating 
an oblique modified shock (with a finite width). In a simplified manner, altering the diffusion conditions 
inside the precursor, this will physically reflect on the magnetic field irregularities within the 
modified shock region. 

\section{Discussion}

It is known that concerning the diffusive shock acceleration mechanism, if one wishes 
to treat more realistically the shock acceleration mechanism itself, one should take into account that 
the shock is not an immaterial surface anymore and the text particle approximation can not hold.
The issue of the back-reaction of the accelerated particles on the shock and the violation 
of the test-particle approximation, occurs when the acceleration process becomes sufficiently 
efficient to generate pressure of the accelerated particles, which is comparable with the incoming 
gas kinetic pressure. The spectrum of the cosmic rays and the structure of the shock are changed by 
this effect which is intrinsically nonlinear. As it is mentioned in the classic review of 
\cite{Jones90}, one could classify the non-linear phenomena in different processes which could 
act on the shock region, considering that the acceleration is efficient. 
One process, is the one where the accelerated particles
can interact with the unshocked plasma creating instabilities, such as different types of plasma 
waves, \cite{Wentzel74}.
Another process, is the one where the particles built up a sufficient pressure 
$P=4\pi/3\int p^3 V f(p,t) dp$, 
which could slow the background plasma before the shock and modify it to an extended structure. 
One can think that as the diffusion coefficient of the particle is increasing with energy, then the very 
high energy particles will escape (having large diffusion lengths) the acceleration region, with a 
considerable energy and momentum. 
This in turn will allow for the compression ratio to 
increase, which will be followed by an increase in the acceleration efficiency. This condition will 
flatten the final spectrum and it will give the opportunity to more particles to get accelerated to 
higher energies, before leaving the system. 
Furthermore, if relativistic particles carry with them most of the energy and produce a significant
part of the pressure in the shock system, then the shock will become cosmic ray dominated and the index 
of the specific heats will decrease due to the relativistic particle speeds. This fact will initiate a further  increase of the compression ratio, making the shock even more efficient, pushing the 
energy towards the very high energy relativistic particles. Nevertheless, an increase in the 
compression ratio is not as easy, as \cite{Ellison91} has shown with simulations and for 
non-relativistic shocks the compression ratio tends to a value of four, being a reasonable 
choice for studies even in the full non-linear case.

\section{Numerical method - Results}


In our present simplified approach, this non-linear problem will be dealt with in a linear way. 
The Monte Carlo numerical approach simulates as realistically as possible
the random process of a physical process; namely in our case, the random walk, the diffusion and the 
scattering of particles, with the supposedly magnetic field irregularities, which can be prescribed 
within the simulation reflecting directly by the different diffusion coefficients.
\begin{figure}[t]
\begin{center}
\includegraphics [width=5.1cm,angle=270]{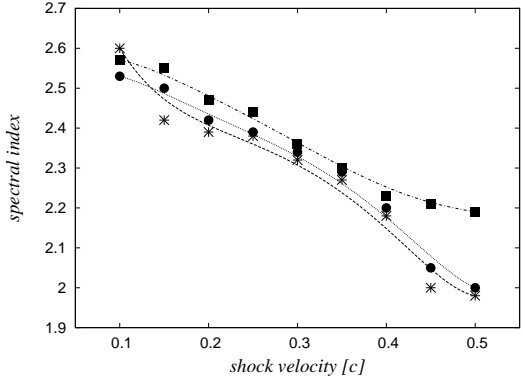}
\end{center}
\caption{The shock velocity versus the obtained 
spectral index for three different simulation runs for a shock inclination equal to 75 degrees. The measurements are taken in the downstream side of the shock rest frame. Starting 
from the top, the first line is for a shock with no precursor, the second is for a shock with a precursor 
width equal to $10\cdot\lambda$  and the third one of precursor width equal to $100\cdot\lambda$. 
One can see that the modified shocks create a flatter spectra compared to the spectrum of sharp shock discontinuity. Also, one may see that the spectral index flattens as the shock's velocity increases. We get similar results for other inclination angles and as long as the shocks are oblique.}
\end{figure}
\begin{figure}[t]
\begin{center}
\includegraphics [width=5.1cm,angle=270]{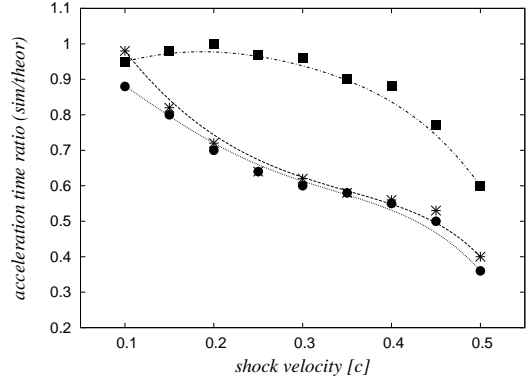}
\end{center}
\caption{Here we show the acceleration time ratio (simulation time / theoretical time) 
versus the shock velocity for a representative shock inclination of 
75 degrees. We get similar results for other angles and as long as the shocks are oblique. 
The measurements are taken in the downstream side of the shock rest frame. Three 
runs have been performed. Starting from the bottom, the first line is for a no precursor shock, the 
second with a precursor width equal to $10\cdot\lambda$ and the third with a precursor width equal to $100\cdot\lambda$. As one sees there is generally a reduction of the acceleration time as higher 
the speed of the shock, in any case, following the same trend of the relativistic shock acceleration 
simulations of past works.}
\end{figure}
A specific feature of the simulations is (following~\cite{Berezhko86}) that there is scattering inside the
precursor (different from what is upstream and downstream of the shock formation)
of  length of $L=\kappa/V$, where $\kappa$ is the spatial diffusion coefficient given by: 
$\kappa= \kappa_{||} cos^2 \psi + \kappa_{\perp} sin^2 \psi$
and $V$ is the velocity of the shock. The angle $\psi$ is the inclination between 
the magnetic field lines and the normal of the precursor.  For parallel shocks one will have
$L=\kappa_{||}/V$, while for perpendicular ones, $L=\kappa_{\perp}/V$, while for all other cases it is $L=\kappa/V$.
In this work we simulate shocks of a high inclination of 75 degrees (as shown in figures 1 and 2), other 
high inclinations are comparable, since as we presented in \cite{MeliBie06}, even for the test-particle approximation, highly oblique shocks can be 
very efficient cosmic ray accelerators, achieving energies of $\sim 10^{17}$ eV.  For a complete picture, 
in a forthcoming detailed paper we present a variety of shock inclinations, ranging from nearly 
parallel to perpendicular ones.
Furthermore, since the cosmic ray diffusion tensor is diagonal in view of the adapted magnetic field 
orientation, one may say that in a Cartesian system $xyz$, $\kappa_{xx} = \kappa_{||}$ and $\kappa_{yy} = \kappa_{zz} = \kappa_{\perp}$, with $\kappa_{||}, \kappa_{\perp}$ the cosmic ray diffusion coefficients longitudinal and transverse relative to the magnetic field.\\
For the simulations a Monte Carlo technique is used for the scattering of the particles in the upstream, 
downstream and precursor media. The downstream spatial boundary required can be estimated from the solution of the convection-diffusion equation in a non-relativistic, large-angle scattering approximation in the 
downstream plasma which gives the chance of return to the shock, $exp(-V_{2}r_{b}/d)$. The mean free 
path is calculated in the respective fluid rest frames (upstream or downstream),
assuming a momentum dependence to this mean free path for scattering along the field,  related to 
the spatial diffusion coefficient $\kappa$, as we mentioned before. 
In the simulation the area of the precursor has a lower diffusion coefficient (where  $\kappa_{||} >> \kappa_{\perp}$ is used (following Berezhko 1987) which physically reflects on an environment 
of a high density and large instabilities.The basic point in this work is that by changing the diffusion conditions (by changing the values of $\kappa$) within the precursor of a width of many times the mean free path, one can study the scattering of the particles which will physically reflect on the irregularity of the magnetic field.
One important point to mention is that the adiabatic invariant is conserved within the precursor since in 
the de Hofmann-Teller frame the adiabatic invariant is conserved anyway by crossing the shock from upstream to downstream and vice versa. Due to this condition, one can calculate a new pitch angle and velocity of the 
particle and the diffusion is continuing until the particle exits the precursor and consequently the 
shock region. Moreover, on the scattering of the particle inside the precursor, we clarify two things: 
The mean free path $\lambda$ is dependent on the distance $d$ inside the precursor in the following 
ways: 1) the diffusion coefficient $\kappa$ is a function of the position and 2) the mean free path is a function of energy. This means that as the particle moves in the precursor, its energy will change and 
hence its mean free path will change as well. The probability that a particle will scatter after moving a distance $\Delta d$, is $\Delta d /\lambda$. Comparing this ratio with a random number, defines as if the
particle will continue to scatter inside the precursor or exit this region downstream or upstream.\\
In figure 1 we show the shock velocity versus the obtained spectral index for three different simulation
runs for a shock inclination equal to 75 degrees. The first is for a shock with no precursor, the second 
is a shock with a precursor width equal to $10\cdot\lambda$  and the third one with a a precursor width 
equal to $100\cdot\lambda$. 
One can see that the modified shocks create flatter spectra which at the 100 mean free paths width,
is flatter then the sharp shock discontinuity and the one with 10 mean free paths width. Also, we may see 
that all spectral indexes flatten as the shock velocity increases. 
In figure 2 the acceleration rate versus the spectral index is shown for a shock inclination of 75 degrees. 
Three runs have been performed. The first  with no precursor, the second with  a precursor width equal 
to $10\cdot\lambda$ and the third with a precursor width equal to $100\cdot\lambda$. As one sees there is
a reduction of the acceleration time. Furthermore, there is a trend for a reduction of the acceleration
time as higher the speed of the shock, making it in any case an efficient cosmic ray accelerator.

\section{Conclusions}

We have preliminarily performed numerical simulations, studying the behaviour of the accelerating 
cosmic rays at highly oblique modified shocks, simulating the percursor with a length 
scale, $L=\kappa/V$, corresponding to the diffusive scale of the energetic particles. 
It is clear that these modified shocks result in flatter spectra compared to the pressumed sharp 
discontinuity shock fronts. This first finding is in accordance with theoretical predictions, since the 
scattering inside the precursor confines high energy particles to further scattering, resulting in higher 
energies making the whole process more efficient. Moreover, one may see  that the spectral index flattens
as the shock velocity increases, leading to spectral index values of those
compared to relativistic shocks. It also seems that even with modified shock fronts, there is 
an acceleration speed up of the process as higher the speed of the modified shock (for speeds $>0.3c$) 
and seem to be as efficient as the relativistic ones, presented by \cite{MeliQa}, \cite{MeliQb}. Additionally, this work has further implications connecting to the results of \cite{MeliBie06}, where it 
is shown that even for the test-particle approximation, highly oblique shocks can be 
very efficient cosmic ray accelerators, achieving energies of $\sim 10^{17}$ eV. There 
is under way a detailed presentation of this work.

\section{Acknowledgements}
The project is funded by the European Social Fund and National Resources (EPEAEK II) PYTHAGORAS.

\bibliography{libros}
\bibliographystyle{plain}
\end{document}